# ENHANCEMENT OF EXCITON EMISSION FROM ZnO NANOCRYSTALLINE FILMS BY PULSED LASER ANNEALING

I. Ozerov[1], M. Arab[1], V.I. Safarov[1], W. Marine*[1], S. Giorgio[2], M. Sentis[3] and L. Nanai[4]

[1] GPEC, UMR 6631 CNRS, Faculté des Sciences de Luminy, Case 901, 13288 Marseille Cedex 9, France.

[2] CRMC2, UPR 7251 CNRS, Faculté des Sciences de Luminy, Case 913, 13288 Marseille Cedex 9, France.

[3] LP3, FRE 2165 CNRS, Faculté des Sciences de Luminy, Case 917, 13288 Marseille Cedex 9, France.

[4] JGYTF, Department of Physics, University of Szeged, Boldogasszony sgt. 6, 6720 Szeged, Hungary.

**Abstract**

Pulsed ArF laser annealing in air and in hydrogen atmosphere improves the optical properties of ZnO nanostructured films. Independently on the ambient atmosphere, laser annealing produces two major effects on the photoluminescence (PL) spectra: first, the efficiency of the exciton PL increases due to decrease of the number of non-radiative recombination centers; second, the intensity of the defect-related orange band decreases because of the removing of excessive oxygen trapped into the films during deposition. However, annealing in the ambient air also increases the intensity of the green band related to oxygen vacancies. We show that the combination of laser annealing and passivation of oxygen vacancies by hydrogen results in films free of defect-related emission and keeps intact their nanostructural character.

* Corresponding author: Tel.: +33 491 829 173; Fax: +33 491 829 176;
E-mail: marine@gpec.univ.mrs.fr (W. Marine)

1. Introduction

Zinc oxide is a very perspective material for applications in optoelectronics. The wide direct gap of 3.37 eV and the large exciton binding energy of 60 meV suggest the use of ZnO in short-wavelength emitting devices. While the optical properties of the bulk material [1] and the epitaxial thin films [2] have been widely studied, the properties of nanostructured ZnO have not been well understood yet. Recently, laser action at room temperature was reported from polycrystalline ZnO films [3]. The laser cavities, assuring the coherent feedback, are formed by multiple scattering of light at grain boundaries. The ZnO grains act as light scattering, amplifying and emitting media. Theoretical predictions suggest the largest oscillator strength of radiative transitions for clusters with sizes in range from 10 to 30 nm because of the strongest coupling of the light with the electronic states in nanoclusters [4]. However, reducing of the size of crystallites to the scale of 10 nm causes a dramatic increase of the ratio of surface and volume. For such small nanoclusters, their physical and chemical properties become strongly influenced by surface quality. In fact, the surface can contain unsaturated dangling bonds, as well as chemisorbed and physisorbed atoms, which modify the light emission from nanoclusters.

Recently, we have proposed the method of synthesis of nanocluster in gas phase using the pulsed laser deposition (PLD) in mixed gas atmosphere [5]. Despite the precautions during the deposition of the films, they contain a considerable concentration of structural defects, especially, the films prepared at low substrate temperature. Usually, nanostructured materials require a special treatment of the surface, reinforcing the stability of the nanoclusters and reducing the possible undesirable influence of the surface. The methods of conventional thermal annealing and surface passivation require heating the samples to high temperature for sufficiently long times for thermal diffusion of the defects to occur. However, the heating of the entire sample has undesirable effects for device production. Using of a pulsed laser is therefore more advantageous for the annealing because laser heating is local and very rapid. The shape and lateral dimensions of the zone affected by laser irradiation can be easily chosen by apertures.

In this paper, we propose a new method of nanocluster surface treatment, combining both annealing of the defects and chemical passivation – pulsed laser annealing in hydrogen atmosphere. This method permits to increase the exciton emission efficiency and to suppress completely the emission related to defects in nanostructured ZnO films keeping intact their morphology.

**2. Experimental**

The nanocrystalline films were prepared by pulsed laser ablation of a sintered ZnO target. The target was placed in a stainless steel chamber on a rotating holder. The quartz glass substrate was placed in front of the target on a holder, equipped with a heater. After evacuating the chamber down to a pressure of about $2\times10^{-7}$ mbar by turbomolecular pump, a continuous flux of oxygen and helium, used as ambient gases, was introduced into the chamber at controlled partial pressures of 4 mbar and 1.5 mbar, respectively. The laser beam was focused onto the target with an incident angle of $45°$ and the fluence was 3.5 J/cm$^2$. In choosing these deposition conditions we were guided by our previous results on optimization of deposition conditions [5]. An ArF excimer laser (Lambda Physik LPX 205) operating at wavelength of 193 nm with pulse duration of 15 ns (FWHM) was used for both, the film deposition and the laser annealing.

The films deposited by pulsed laser ablation were homogenous, transparent and colorless. We estimated the thickness of the films to be about 1 μm from light interference fringes. Only the flat central part of samples with area of about 2 cm² was studied to avoid the possible effects of films thickness on their properties. In the annealing experiments, the most homogenous part of laser beam was chosen by apertures and focused on a spot of 18 mm² on the sample surface at normal incidence. The samples were placed in a chamber under controlled atmosphere of air or hydrogen. For films characterization, the same laser spot was used during the studies of the effect of the number of shots per site at fixed fluence. For the annealing with different laser fluences, every series of shots used a fresh spot.

The crystallographic structure and orientation of the films were analyzed by x-rays diffraction (XRD). The surface morphology of the films before and after annealing was examined by atomic force microscope (AFM) operating in tapping mode. The size and shape of the nanoclusters were observed by high resolution transmission electron microscopy (HRTEM).

The photoluminescence of the films at room temperature was excited by a 100 W mercury lamp ($\lambda = 254$ nm). The spectra were detected by a cooled photomultiplier (Hamamatsu R943-02) coupled to a grating monochromator.

## 3. Results and discussions

The XRD results showed that nanostructured films crystallize in the wurtzite structure with c-axis perpendicular to the substrate surface. The conventional PLD of c-axis oriented polycrystalline ZnO films onto fused silica substrates was reported earlier [6]. In this case the films are grown from atomic species arriving onto the substrate where the film growth and crystallization take place. The main difference of our deposition procedure from the one reported in [6] is the relatively high ambient gas pressure allowing the crystallization of nanoclusters in gas phase before their deposition. Thus, a quite dense nanostructured film is formed using the separate clusters as building blocks. Even in the case of amorphous substrates, the c-axis orientation of the films is favorable energetically because of polarity of oxygen- and zinc-terminated (0001) atomic planes of wurtzite structure. The interaction between the polar surfaces of ZnO nanocrystallites with oxide substrate surface is probably responsible for the oriented deposition at the beginning of the film growth. The following deposition keeps the c-axis orientation of previously deposited cluster layers.

HRTEM image and the corresponding diffraction pattern of a sample obtained before the laser treatment (Fig. 1) show that the nanocrystalline films are consisted of near-spherical clusters with an average size of about 8 nm. The atomic plane spacing obtained from diffraction pattern corresponds to the wurtzite phase of ZnO confirming the results obtained by XRD. Neither the coalescence nor strong chemical bonding between the clusters was observed in as-prepared films.

During laser annealing, the photons with energy of 6.4 eV can produce two major effects on the film surface: thermal and photochemical. Laser irradiation induces a rapid local increase in temperature that provokes the diffusion of electronic defects to the surface or along grain boundaries. The weakening of the chemical bonds can result in preferential ejection or evaporation of certain atomic or ionic species as well as in annihilation of pairs of defects like an interstitial and a vacancy. The photochemical action of the laser light is manifested by: (i) the desorption of atoms from the surface, chemically or physically adsorbed during the film deposition or the exposition to the atmosphere; (ii) the displacement of atoms from their regular positions in crystalline lattice, and, hence, the formation of structural defects in the crystalline structure; (iii) the excitation and photolysis of the molecules present in annealing ambient atmosphere.

Both thermal and photochemical effects of laser irradiation strongly depend on laser fluence. At low laser fluences, when the melting was not achieved, pulsed laser annealing kept intact the structural properties of nanostructured films. The results of microscopic analysis are similar to those obtained on as-prepared films. Contrarily, the optical properties of films change drastically.

Figure 2 shows the typical PL spectra of a film deposited on a $SiO_2$ substrate at room temperature and annealed in ambient air with low laser fluence of 75 mJ/cm². The spectra consist of three bands. The first one, centred at photon energy of 3.28 eV, corresponds to the radiative recombination of excitons. The second and the third PL bands, related to deep defect levels, overlap and were fitted for all the spectra by two Gaussian lines (cf. Fig. 2) centered at 2.00 eV (with spectral width of 0.64 eV FWHM) and at 2.40 eV (width of 0.61 eV FWHM) for orange and green bands, respectively. The orange band is due to the regions with local oxygen excess [1, 5]. The green band corresponds to the PL of oxygen vacancies [1,7]. Figure 3a shows the evolution of the integral intensities of the two defect bands for different numbers of laser shots per site. The decrease in intensity of the orange band corresponds to release of excessive oxygen, which was trapped into the films during deposition in the oxygen-reach atmosphere [5]. The accompanying increase of the green band intensity is explained by creation of oxygen vacancies in the films, this band being due to radiative recombination of photogenerated hole with electron at the level of a singly ionized oxygen vacancy [7].

Figure 3b shows the evolution of intensities of exciton emission and green defect band as a function of the number of laser shots per site at a fluence of 120 mJ/cm², just below the melting threshold. It can be clearly seen that the intensity of the exciton peak increases rapidly at the beginning and saturates at about 5 shots per site. This result confirms our suggestion that the laser annealing decreases the concentration of non-radiative recombination centers on the surface of the nanoclusters. To obtain the same rise in intensity of exciton band by annealing at lower fluences more laser shots were needed. During annealing with multiple shots the increase in intensity of the exciton band is accompanied by increase of the green band because of creation of oxygen vacancies. The annealing with 3 laser shots at fluence of 120 mJ/cm² is found to be optimal because the intensity of exciton peak and the ratio of exciton and green band intensities reach the maximum. In spite of increase of the optical quality of ZnO nanostructured films (the enhanced exciton emission, the complete suppression of the orange band) laser annealing in ambient air raises also the green PL band which should be passivated.

The influence of the chemical passivation by hydrogen on optical properties of ZnO bulk crystals was recently reported [8]. The hydrogenation by plasma treatment increased the efficiency of the band-edge emission. The main problem is that molecular hydrogen is relatively inert at normal conditions and the passivation requires the special conditions of hydrogen insertion into ZnO. Such treatment can be incompatible with the desired optical properties and the nanocrystalline structure required for realization of a random laser. For example, hydrogen ion-implantation in ZnO reduces the intensity of the exciton emission from ZnO because the defects, acting as non-radiative recombination centers, are induced by ion beam [9]. The formation of OH complexes in ZnO single crystals was observed by infrared absorption spectroscopy [10]. The thermal annealing at very high temperature (700°C) during several hours was necessary for these complexes to be observed.

The excitation and photolysis of hydrogen molecules by ArF laser in a region close to the surface of the film allows the chemical passivation without long heating or bombarding of the samples with energetic ions. The dissociation energy of $H_2$ molecule is 4.2 eV [11], and the photons with energy of 6.4 eV can efficiently photodissociate hydrogen molecules. The produced hydrogen atoms in excited state are more reactive chemically and form chemical bonds on the surface of ZnO nanoclusters.

The PL spectra of a film prepared at low substrate temperature (220°C) and annealed with the same fluence in air and in $H_2$ are shown in Figure 4. The photoluminescence of the as-prepared film is shown in Fig. 4a for comparison. The exciton peak is well pronounced, but its intensity is reduced. The orange band centered at 2.0 eV is also present in the film. After annealing in air, the orange band disappeared and the green band increased in intensity (Fig. 4b). An annealing of the sample with the same laser fluence in hydrogen atmosphere leads to a suppression of both defect-related luminescence bands (Fig. 4c). The intensity of exciton band increases about four times compared with as-prepared sample, after laser treatment in both air and hydrogen.

The absence of the green band in PL spectra of the films annealed in hydrogen environment suggests that hydrogen atoms participate in the formation of complex defects containing oxygen vacancies, which do not give raise to the green luminescence. One of the possible complex defects is a substitutional hydrogen atom on an oxygen site. To the best of our knowledge, the experimental confirmation of such type of hydrogen configuration was never reported before. In addition, recent theoretical investigations [12] predicted the possibility of the formation of such a complex and suggested its behavior as a shallow donor. Our preliminary measurements of the electrical conductivity show that a treatment in hydrogen atmosphere decreases the resistance of the films. However, the electron transport in an assembly of ZnO nanoclusters was found to be critically determined by the quantum properties of the building blocks [13] and more detailed study is needed. We suggest that hydrogen occupies oxygen vacancies and quenches the green emission considerably enhancing the optical properties of ZnO nanocrystalline media.

If the laser fluence is high enough (more than 130 mJ/cm²), the irradiated surface melts and resolidifies after the end of the laser pulse. The fact that laser heating of the nanocrystalline films up to melting strongly changes their microstructure is evidenced from the observations by HRTEM. Figure 5 shows the image of the same film as in Fig. 1 after annealing with laser fluence of 140 mJ/cm². The nanoclusters are more compactly packed in the film. Moreover, the clusters grow up to sizes of 40-60 nm by coalescence always keeping the wurtzite crystalline structure. The faceting and the boundaries of the crystallites can be clearly seen in Fig. 5. In addition to microscopic changes, the surface morphology of the film at mesoscopic scale was also strongly modified. The typical dimensions of the agglomerations of nanoclusters observed at AFM images (Fig. 6) are roughly the same (300-500 nm) both before and after annealing. The scale of height of both images is the same, so the quantity of the material on the surface observed by AFM is the same. However, the presence of interconnected grains in form of islands surrounded by holes after annealing shows the increase in nanocluster density in the agglomerations. So, both microscopic methods present the evidence of the film melting.

The PL spectra of the films treated with high laser fluences show the green band independently on the ambient atmosphere. Once the green luminescent band appeared, it remained very stable and could not be suppressed by following treatment.

## 4. Conclusions

We have demonstrated that UV pulsed laser annealing of nanostructured ZnO films with fluences below the melting threshold drastically improves their optical properties keeping intact their morphology. The intensity of the exciton emission strongly increases during the laser annealing. For all laser fluences after several laser shots the saturation of the exciton peak emission intensity was observed principally due to the suppression of the majority of the non-radiative centers as a result of the annealing. Increasing laser fluence or/and number of shots per site allows one to remove the orange PL band originated from the defects related to local oxygen excess. After the annealing of the films in the air, the green band related to oxygen vacancies appears. The annealing in hydrogen with low laser fluences prevents the appearance of this band. We believe that complexes of oxygen vacancy with hydrogen are formed. Being shallow donors, these complexes do not contribute to the non-radiative recombination of excitons and completely suppress the green emission. Laser annealing with fluences up to those needed for melting increases the exciton emission intensity but also leads to changes in the morphology of the films, nanocluster coalescence and in the formation of oxygen-deficient films.

Reference number of Manuscript: F-III.4

List of Graphics Captions

Fig. 1. HRTEM image of ZnO nanoclusters. The inset presents the corresponding electron diffraction pattern.

Fig. 2. Room temperature photoluminescence spectra of ZnO nanocrystalline film annealed in ambient air at laser fluence of 75 mJ/cm²: (a) 30 laser shots; (b) 120 shots. The dotted lines present the Gaussian fitting.

Fig. 3. PL integral intensity vs. number of laser shots per site: (a) Fluence $F = 75$ mJ/cm²; (b) $F = 120$ mJ/cm². (Exciton peak – triangles; Green band – circles; Orange band – squares).

Fig. 4. Room temperature photoluminescence spectra of ZnO nanocrystalline film. (a) as-prepared film; (b) annealed in ambient air by 3 shots of ArF laser with the fluence $F = 120$ mJ/cm²; (c) annealed in hydrogen by 3 shots of ArF laser with the fluence $F = 120$ mJ/cm².

Fig. 5. HRTEM image of ZnO nanoclusters after laser annealing with fluence $F = 140$ mJ/cm².

Fig. 6. AFM image of ZnO film: (a) as-deposited; (b) annealed at laser fluence $F = 140$ mJ/cm².

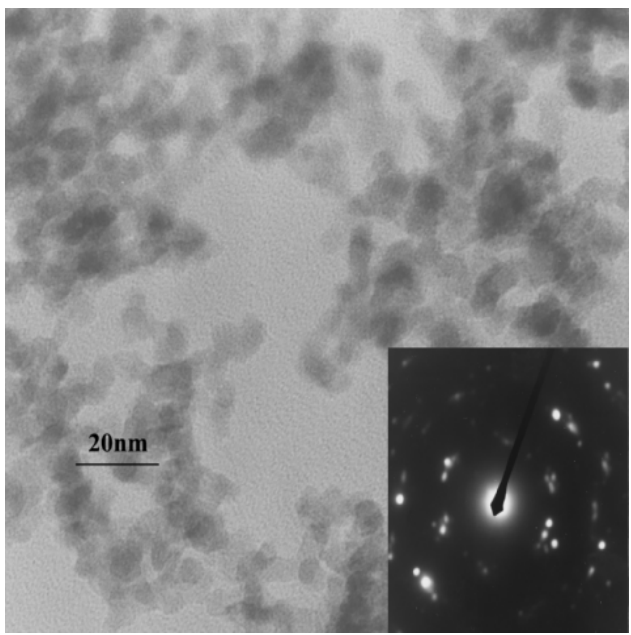

Fig. 1
Ozerov et al. - E-MRS 2003 - F-III.4

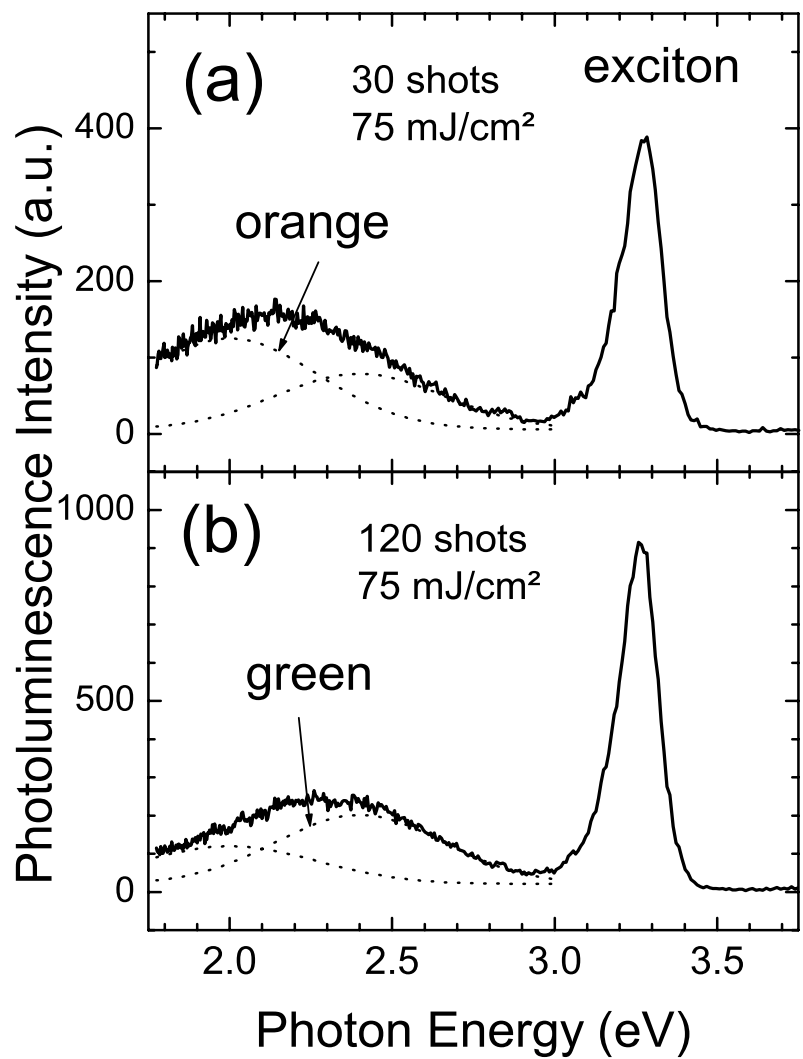

Fig. 2
Ozerov et al. - E-MRS 2003 - F-III.4

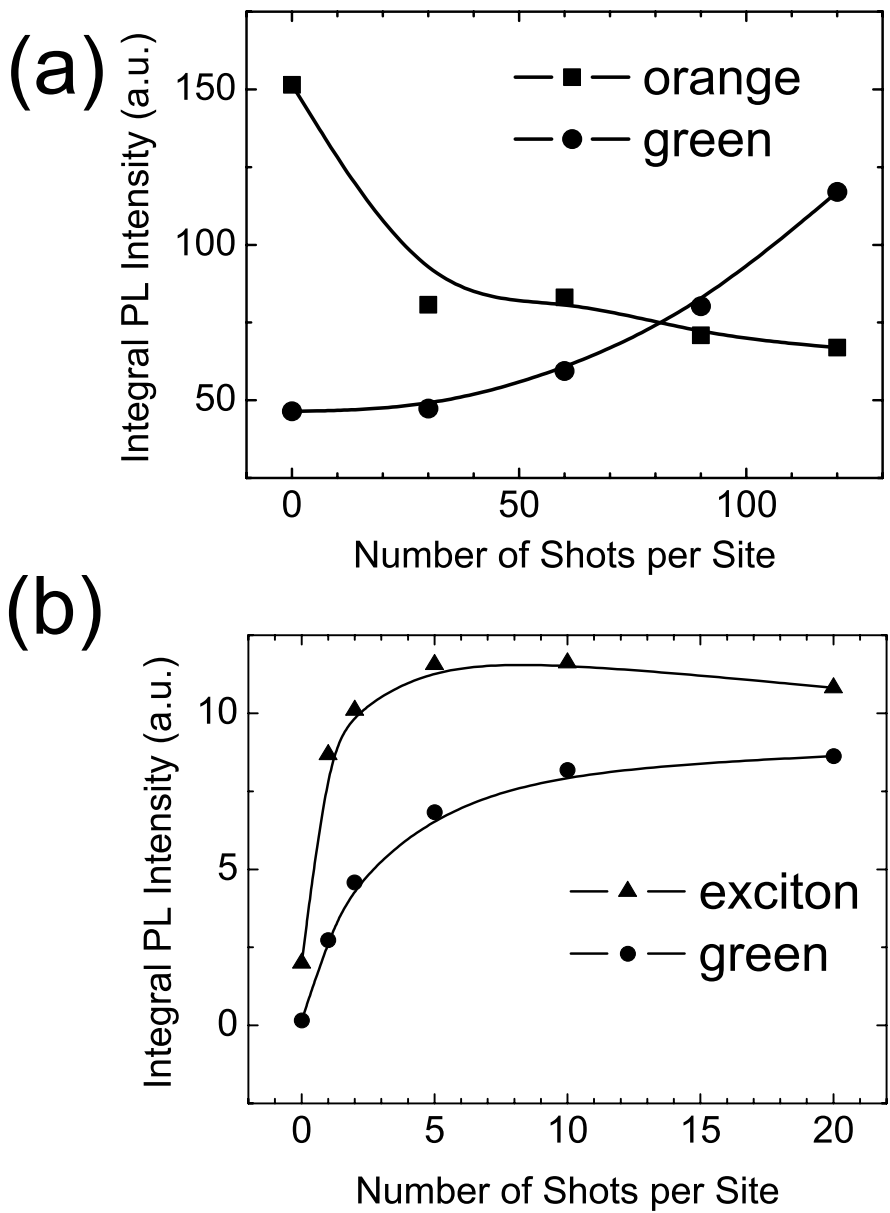

Fig. 3
Ozerov et al. - E-MRS 2003 - F-III.4

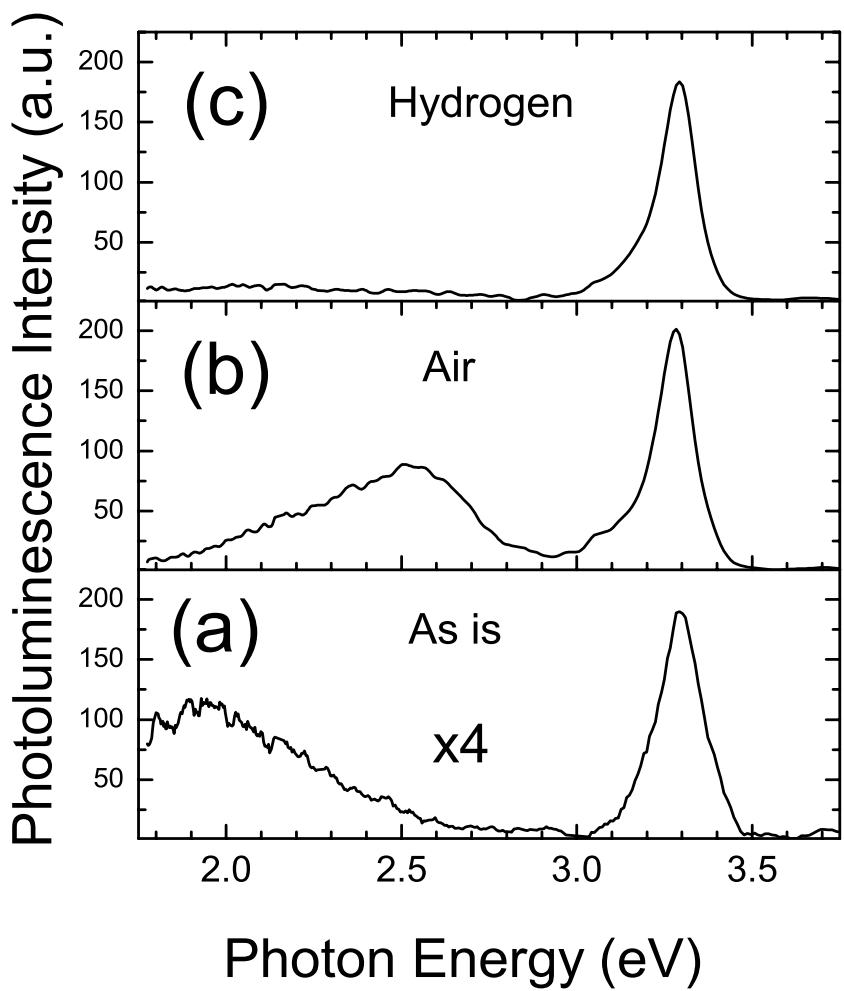

Fig. 4
Ozerov et al. - E-MRS 2003 - F-III.4

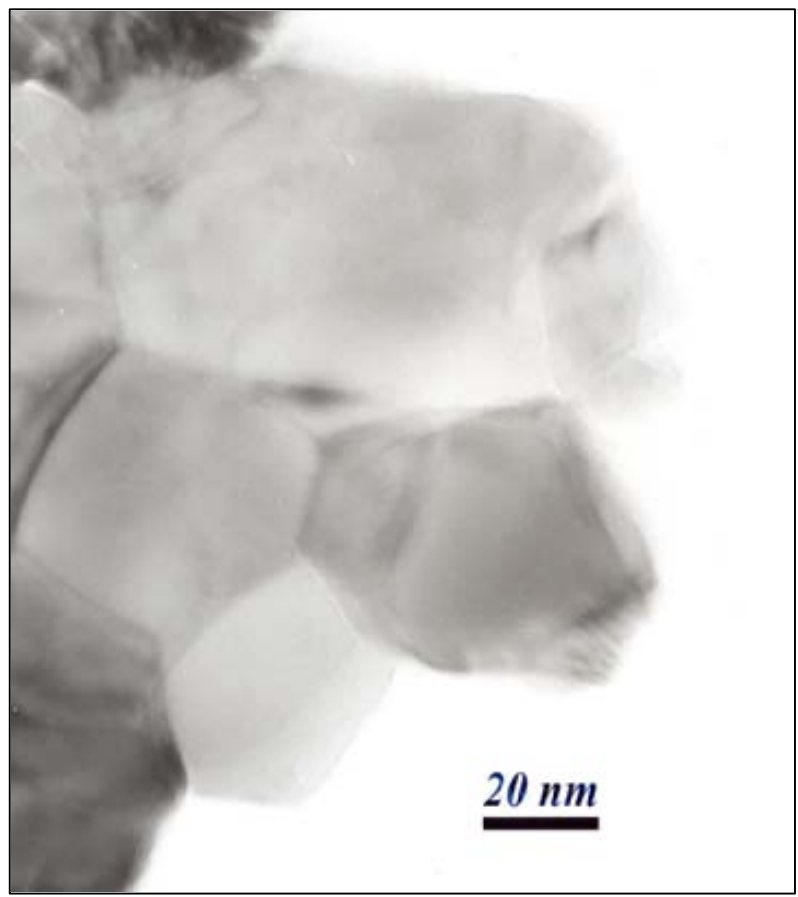

Fig. 5
Ozerov et al. - E-MRS 2003 - F-III.4

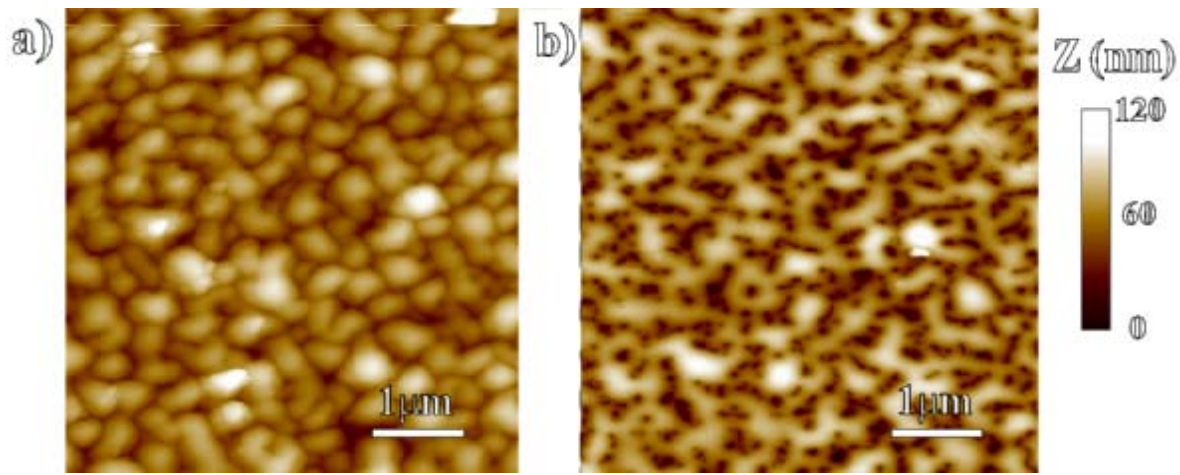

Fig. 6
Ozerov et al. - E-MRS 2003 - F-III.4